\newcount\style \style=2
\ifodd\style
 \documentstyle[aas2pp4,epsf]{article}
  \else
   \documentstyle[12pt,aasms4,epsf]{article}
\fi %\input psfig

\begin{document}
\setlength{\baselineskip}{12pt}

\newcommand\bb[1] {   \mbox{\boldmath{$#1$}}  }

\newcommand\vv {\bb{v}}
\newcommand\ve {\bb{v_e}}
\newcommand\B {\bb{B}}
\newcommand\del{\bb{\nabla}}
\newcommand\bcdot{\bb{\cdot}}
\newcommand\btimes{\bb{\times}}
\newcommand\BV{Brunt-V\"ais\"al\"a\ }
\newcommand \Ombis{  {d\Omega^2\over dR}  }
\newcommand \lnOmbis{  {d\ln \Omega^2\over d \ln R}  }

    \def\dd{\partial}
    \def\tilde{\widetilde}
    \def\etal{et al.}
    \def\eg{e.g. }
    \def\etc{{\it etc.}}
    \def\ie{i.e.}
    \def\beq{ \begin{equation} }
    \def\eeq{ \end{equation} }
    \def\spose#1{\hbox to 0pt{#1\hss}} % from Scott Tremaine
    \def\ltsim{\mathrel{\spose{\lower.5ex\hbox{$\mathchar"218$}}
	 \raise.4ex\hbox{$\mathchar"13C$}}}

\def\tilde{\widetilde}

\long\def\Ignore#1{\relax}

\title{Linear Analysis of the Hall Effect in Protostellar Disks}

\author{Steven A.~Balbus\altaffilmark{1}}
\altaffiltext{1}{Virginia Institute of Theoretical Astronomy, Dept.~of
Astronomy, University of Virginia, Charlottesville, VA 22903-0818 --- sb@virginia.edu}
%\affil{Virginia Institute of Theoretical Astronomy,}
%\affil{Department of Astronomy,}
%\affil{University of Virginia,}
%\affil{Charlottesville, VA 22903-0818}
%\affil{sb@virginia.edu}

\author{Caroline Terquem\altaffilmark{2,3}}
\altaffiltext{2}{Institut d'Astrophysique de Paris, 98 bis Blvd.~Arago, 75014
Paris, France --- terquem@iap.fr}
\altaffiltext{3}{Universit\'e Denis Diderot--Paris VII, 2 Place Jussieu, 75251 Paris
Cedex 5, France}
%\affil{Institut d'Astrophysique de Paris,}
%\affil{98 bis Blvd. Arago}
%\affil{Paris 75014, France}
%\affil{terquem@iap.fr}

\vskip 2 truein
\centerline{\today}
%\centerline{To appear in the {\sl Astrophysical Journal, {\bf 521,}} 20 August 1999.}
\vskip 2 truein

\begin{abstract}

The effects of Hall electromotive forces (HEMFs) on the linear
stability of protostellar disks are examined.  Earlier work on this
topic focused on axial field and perturbation wavenumbers.  Here we
treat the problem more generally.  Both axisymmetric and
nonaxisymmetric cases are investigated.  Though seldom explicitly included
in calculations, HEMFs appear to be important whenever Ohmic
dissipation is.  They allow for the appearance of electron whistler
waves, and since these have right-handed polarization, a helicity
factor is introduced into the stability problem.  This factor is
the product of the components of the angular velocity and magnetic
field along the perturbation wavenumber, and it is destabilizing when
negative.  Unless the field and angular velocity are exactly aligned,
it is always possible to find destabilizing wavenumbers.  HEMFs can
destabilize any differential rotation law, even those with angular
velocity increasing outward.  Regardless of the sign of the angular
velocity gradient, the maximum growth rate is always given in magnitude
by the local Oort A value of the disk, as in the standard
magnetorotational instability.   The role of Hall EMFs may prove
crucial to understanding how turbulence is maintained in the ``low
state'' of eruptive disk systems.


\end{abstract}

\section{Introduction}

The stability of differentially rotating gas disks depends sensitively
upon whether or not a magnetic field is present.  A weak
(subthermal) magnetic field undermines the stabilizing influence of
Coriolis forces (Balbus \& Hawley 1991), and the resulting turbulence
greatly enhances internal angular momentum transport (Hawley, Gammie,
\& Balbus 1995; Balbus \& Hawley 1998).  This important behavior, which 
we refer to here as the standard magnetorotational instability (MRI),
probably is the underlying cause of ``anomalous viscosity'' in
accretion disks.  While other modes of transport (e.g.~waves) are
certainly possible, the classical enhanced {\em turbulent} transport
associated with $\alpha$ disks is almost surely MHD turbulence (Balbus
\& Papaloizou 1999).

Protostellar disks remain problematic.  They seem too dense and too
cool to be magnetically well-coupled over their full radial extent.
Understanding how protostellar and protoplanetary disks behave is an
ongoing theoretical challenge.  It has also been recently noted (Gammie
\& Menou 1998, Menou 2000) that the outer regions of dwarf novae disks
may have a very low ionization fractions.  That protostellar disks and
dwarf novae both show episodic eruptions may be no coincidence; in both
systems the question of magnetic coupling (and therefore the presence
of MHD turbulence) is a delicate one.  This paper addresses a very
small but fundamental part of this problem: the linear effects of Hall
electromotive forces (HEMFs) on disk dynamics.  While the results are
more broadly applicable, protostellar disks will be the focus of this
paper.

The importance of the Hall effect in modifying the MRI was pointed out
in a recent paper by Wardle (1999; hereafter W99), who analyzed the
linear stability of a protostellar disk in the presence of a vertical
magnetic field.  Such disks are generally in the Hall regime.  That is,
their combination of densities, temperatures and ionization fractions
places them in a realm of parameter space in which departures from
ideal MHD are important, which are strongly influenced by HEMFs.  The
Hall effect tends to be important simultaneously with finite
conductivity.  Numerical simulations of low ionization disks (e.g.,
Fleming, Stone, \& Hawley 2000) have not yet included Hall terms, but
have focused rather exclusively on finite conductivity.

W99 found that standard, locally unstable MRI modes may either be
stabilized or destabilized depending upon whether $\bb{\Omega\cdot B}$
is positive or negative.  (Here, as usual, $\bb{\Omega}$ is the local
angular velocity vector of the disk, and $\bb{B}$ is the magnetic
field.)  In other words, HEMFs appear to introduce a helicity
effect into the problem.

This immediately raises an interesting question: what happens if
$\bb{\Omega\cdot B} = 0$?  Would the Hall effect disappear?  Since the
W99 analysis was restricted to axial fields and axial wavenumbers, it
could not address this problem.  Clearly HEMFs do not simply vanish
when the magnetic field lies in the disk plane, and the identification
of $\bb { \Omega\cdot B}$ as the fundamental Hall stability parameter
therefore seems suspicious.  Identifying this key parameter more precisely
is important, for it is at the heart of MHD stability in low ionization
disks.

We are motivated, therefore, to examine the stability of Hall regime
accretion under very general circumstances.  We restrict neither the
local field geometry, nor the local wavenumber direction.  Furthermore,
we have chosen to couch the analysis in a dynamical context, instead of
ascribing the inductive Hall terms to effective conductivities (W99).
The effective conductivity approach offers terseness and mathematical
economy; the dynamical approach makes the physics of the coupling
between the magnetic field and the disk coupling more transparent.
Both methods must of course ultimately be equivalent.  It is
enlightening, however, to view this important problem from a different
physical perspective, and we have accordingly chosen this route.

The plan of the paper is as follows.  In \S 2, we discuss the basic
formulation of the problem.  This includes the various physical and
compositional parameters, the fundamental equations, and an order of
magnitude estimate for the relative sizes of the Ohmic, Hall, and
ambipolar diffusion terms in the induction equation.  Section 3
presents the linear dispersion relation for a uniformly rotating disk,
because this is the simplest setting to see the interplay between
rotation and the Hall EMFs.  Section 4 is a detailed and extensive
analysis of differentially rotating disks for arbitrary field
geometries and perturbation wavenumbers.  Section 5 summarizes the
findings of this paper.

\section {Basics}

\subsection {Physical and Compositional Parameters}

A typical protostellar disk consists of molecular gas with cosmic
abundances, and spans a temperature range from 10 K in its outer
regions to a few $10^3$ K near the central star.  The characteristic
disk size may be up to of order 100 AU, and its total mass will
typically be less than roughly 0.1 of the central star's mass.  If 0.01
$M_\odot$ is spread over a cylinder of radius 1 AU and height 0.1 AU,
this would yield an average mass density $\overline{\rho} \simeq
2\times 10^{-10}$ g cm$^{-3}$, and an integrated column density of
$\Sigma \simeq 3\times 10^2$ g cm$^{-2}$, fiducial numbers one should
bear in mind.

The dynamics of the disk is likely to be dominated by the Keplerian orbital velocity
$v_K$:
\beq
v_K^2 = {GM\over R}\equiv R^2 \Omega^2
\eeq
where $M$ is the central mass, $R$ the cylindrical radius, and $\Omega$
the angular velocity.  The disk is assumed to be thin; the
characteristic temperature $T$ is much less than any virial
temperature.  The scale height $h$ (defined below) satisfies $h/R \ll
1$.

The number density of species $X$ is denoted $n_X$.  The H$_2$ and He number densities are
assumed to be related by 
\beq
n_{He} = 0.2 n_{H_2}  ,
\eeq
which implies a total neutral number density 
\beq
n = n_{H_2} + n_{He} = 1.2 n_{H_2},
\eeq
and a neutral mass density 
\beq
\rho = 2.8 n_{H_2} m_p,
\eeq
where $m_p$ is the proton mass.  This gives a mean mass per particle of
\beq
\mu = \rho/n = 2.33 m_p,
\eeq
and isothermal and adiabatic sound speeds respectively of
\beq
c_S^2 = 0.429 k T/m_p, \qquad a^2 = 0.6 kT/m_p,
\eeq
where $k$ is the Boltzmann constant.  For simplicity we will define the disk 
scale height $h$
by $h = c_S/\Omega $.  Under conditions of interest, the dominant ion will
generally be once-ionized potassium K${^+}$, and accordingly we take a mean ion mass of 
\beq
m_i = 39m_p
\eeq
In molecular cloud studies, $m_i=30m_p$ is more typically used (e.g., Draine,
Roberge, \& Dalgarno 1983).  

The coupling between ions and neutrals depends upon the rate coefficient, denoted
$\langle \sigma v\rangle_{iN}$.  In essence, this is the
the product of a cross section with the
component of the relative velocity along the drift axis, averaged over 
the electron Maxwellian distribution (Spizter 1978).  Similarly, there is an 
electron-neutral coupling rate, $\langle \sigma v\rangle_{eN}$.  
We follow Draine et al.~(1983) and Blaes \& Balbus (1994), adopting the values
\beq
\langle \sigma v\rangle_{iN} = 1.9 \times 10^{-9} {\rm cm}^3\, {\rm s}^{-1}  \qquad
\langle \sigma v\rangle_{eN} = 10^{-15} \left( 128kT\over 9\pi m_e \right)^{1/2}= 8.28\times
10^{-10} T^{1/2} {\rm cm}^3\, {\rm s}^{-1}
\eeq
The force per unit volume on the neutral fluid due to ion drag is given by an expression of the
form (Shu 1992)
\beq
\bb{f_{ni}} = - \gamma \rho \rho_i (  \vv - \bb{v_i}),
\eeq
where $\rho_i$ is the ion mass density, $\vv$ the neutral
velocity, $\bb{v_i}$ the ion velocity, and $\gamma$ is the so-called drag coefficient,
\beq
\gamma = {\langle \sigma v\rangle_{iN}\over m_i + \mu}= 2.75 \times 10^{13} {\rm cm}^3\,
{\rm s}^{-1} {\rm g}^{-1}.
\eeq
Our numerical value differs somewhat from Draine et al.~(1983) because we use the potassium
value for the ion mass $m_i$.  

Electron-neutral coupling results in a finite electrical conductivity for the disk gas
(Krall \& Trivelpiece 1973):
\beq
\sigma_e = {n_e e^2\over m_e \nu_{eN}} ={n_e e^2\over m_e n \langle \sigma v\rangle_{eN}}  ,
\eeq
where $e>0$ is the magnitude of the electron charge, $n_e$ is the electron number
density, and $\nu_{eN}$ is the electron-neutral collision frequency. 
The associated resistivity is
\beq
\eta = {c^2\over 4 \pi \sigma_e} = 234 \left(n\over n_e\right)
T^{1/2} {\rm cm}^2 {\rm s}^{-1},
\eeq
where $c$ is the speed of light. 
The dimensionless measure of the relative importance of resistivity is
the magnetic Reynolds number,
\beq
Re_M = {v_A h\over \eta},
\eeq
where $v_A$ is the magnitude of the Alfv\'en velocity
\beq
\bb{v_A} = { \bb{B}\over \sqrt{ 4\pi\rho} },
\eeq
and $\bb{B}$ is the magnetic field vector.

Finally, the cyclotron frequencies of the electrons, ions, and neutrals are denoted
\beq
\omega_{c\alpha} = {eB\over m_\alpha c}
\eeq
where $\alpha$ is respectively $e$, $i$, and $\mu$ ($m_\mu\equiv\mu$).
%Finally, the electron cyclotron and ion cyclotron frequencies are respectively
%\beq
%\omega_{ce}  = {e B\over m_e c} = 1.76 \times 10^7 \, B\ {\rm rad\ s}^{-1} ,
%\qquad \omega_{ci} = {eB\over m_i c} = 246 \, B\,  {\rm rad\ s}^{-1}. 
%\eeq
Although the neutrals are of course not directly affected by the
magnetic field, the neutral cyclotron frequency is an important
characteristic frequency of the analysis.

\subsection{Equations}

The fundamental dynamical equations are
mass conservation,
\begin{equation}\label{mass}
{\dd\rho\over \dd t} + \del\bcdot (\rho \bb {v}) =  0
\end{equation}
the dynamical equation of motion,
\beq\label{mom}
\rho {\dd\vv \over \dd t} + (\rho \vv\bcdot\del)\vv = -\del\left(
P + {B^2\over 8 \pi} \right)-\rho \del \Phi +
\left( {\B\over 4\pi}\bcdot \del\right)\B
\eeq
and the induction equation,
\beq
{\dd\B\over \dd t} = \del\btimes\left( \ve \btimes {\B} -
\eta \del\btimes{\B} \right).
\eeq
Here, $\vv$ is the velocity of the neutrals, and $\ve$ is the electron
velocity, and $\Phi$ is the central
gravitational potential.

In
equation (\ref{mom}), which is dominated by the neutral component,
we have made implicit
use of the low inertia limit for the ions.  Ion momentum balance 
is set by balancing the ion-neutral drag force with
the Lorentz force,
\beq\label{i_mom}
% \gamma\rho\rho_i( \vv - \bb{v_i}) +
{1\over c} \bb{J\times B} = \bb{f_{ni}},
\eeq
since all other terms in the ion momentum equation are small.
It is actually $\bb{f_{ni}}$ that appears on the right hand
side of equation (\ref{mom}); the low ion inertia approximation
consists of instantaneously balancing this force against the Lorentz
force.

\subsection {The Induction Equation}

Note the appearance of the electron velocity $\ve$ in the induction
equation.  In ideal MHD, the distinction between the fluid velocity and
the electron velocity is unimportant.  But in fact, in the absence of
resistance, it is the electron fluid in which the magnetic field is
``frozen,'' since the electrons are the mobile charge carriers.  The
difference between ion and neutral velocities is responsible for
ambipolar diffusion, and the difference between ion and electron
velocities gives rise to the Hall effect (e.g., Wardle \& K\"onigl
1993).

In general,
\beq\label{ve}
\ve = \vv + (\ve - \bb{v_i})+ (\bb{v_i} - \vv) 
= \vv  - {\bb{J}\over n_e e}+{ \bb{J\times B}\over \gamma\rho\rho_i c},
\eeq
We are assuming that any ions present are singly ionized (which is
appropriate for a very low ionization gas), so that the electron and
ion number densities are equal.  In the ideal MHD limit, both $\ve -
\bb{v_i}$ and $\bb{v_i} - \vv$ are assumed to be negligibly small.
Retaining them in the full induction equation leads to

\beq\label{ind}
{\dd\B\over \dd t} = \del\btimes\left( \vv
\btimes {\B} - {4\pi\eta\bb{J}\over c} - { \bb {J\times B}\over n_e e} + {
\bb{ (J\times B)\times B}\over c \gamma\rho_i\rho} \right),  \eeq
where we have used the standard MHD approximation
\beq
\bb{J} = {c\over4\pi} \bb{\del\times B}.
\eeq

The four terms on the right side of the induction equation
can be arranged by order of magnitude.  Reading
from left to right, we denote them $I$ (Inductive), $O$ (Ohmic), $H$
(Hall), and $A$ (Ambipolar).  Assume that typical fluid velocities are of order
$v_A$, and that typical gradients are of order an inverse scale-height $1/h$.
Then, the relative sizes of these terms are
\beq {O\over
I} \sim {1\over  Re_M}, \quad {H\over O} \sim {\omega_{ce} \over
\nu_{en}} , \quad {A\over H} \sim {\omega_{ci}\over\gamma\rho}
\eeq
The first ratio just reflects the familiar fact that when the magnetic
Reynolds number approaches unity, ohmic losses are important.  The
results of \S 2.1 give
\beq
{O\over I} \sim {1\over Re_M} = 4.2\times 10^{-14} \left( n\over n_e\right)
\left( c_S\over v_A\right) \left(1 \>{\rm AU}\over R\right)^{3/2}  
\left( 10^3\over T \right)^{1/2} \left(M\over M_\odot\right)^{1/2},
\eeq
and 
\beq
{H\over O} \sim {\omega_{ce} \over
\nu_{en}} \simeq \left( 8\times 10^{17}\over n\right)^{1/2} \left(v_A\over
c_S\right) , \quad {A\over H} \sim {\omega_{ci}\over\gamma\rho} \simeq
\left( 9\times 10^{12}\over n \right)^{1/2} \left( T\over 10^3\right)^{1/2}
\left(v_A\over c_S\right).  \eeq
Under interstellar conditions $1/Re_M$ is tiny, even though $n/n_e$ may
be large.  (Note that in equation [13], $h$ would be a characteristic
interstellar length, not a disk thickness.)  In this case, neither the
$O$ nor the $H$ term is important.  But the ambipolar diffusion $A$
term may be non-negligible.  This leads to a regime of ideal MHD in the
ions, which then couple the magnetic field to the neutrals via
ion-neutral collisions.

In protostellar disks, our interest here, the far smaller ionization
fraction can reduce $Re_M$ to the point where resistivity can affect
the dynamics.  This Reynolds number need not be of order unity in an
MHD-turbulent fluid before it is significant; some field configurations
are affected when $Re_M$ is larger than $10^4$ (Fleming, Stone, \&
Hawley 2000).  The key point here, however, is not the well-known
result that resistivity is important in protostellar disks, it is the
less appreciated fact that HEMFs will generally be important in a
magnetized disk whenever resistivity is.  The two should be treated on
the same footing (W99).

The importance of the Hall term relative to Ohmic losses is noteworthy
because it is surprisingly general.  As formulated above, for a
given ratio of magnetic to thermal energy,
the relative importance
of the Hall term is {\it independent\/} of the ionization level, and
depends only upon the number density of the dominant neutrals.
Furthermore, the relative importance of the Hall term is not 
a feature unique to gases of low ionization fraction.  Using
a nominal resistivity value of $5\times 10^{12}T^{-3/2}$
cm$^2$ s$^{-1}$ for a
fully ionized plasma (Spitzer 1962), we find
\beq
{H\over O} \sim {\omega_{ce}\over \nu_{ei}} \simeq 4 \left(v_A\over c_S\right) 
\left(T\over 10^5\right)^2 \left(10^{16}\over n\right)^{1/2},
\eeq
where $\nu_{ei}$ is the electron-ion collision rate.  
We see that if Ohmic dissipation is of interest (as it would be for magnetic
reconnection), the temperature and density regimes of ionized accretion disks
imply that the Hall effect cannot be ignored. 

At typical disk protostellar disk densities, ambipolar diffusion can be
neglected, though it may become important in the outermost regions of
the disk.  Since 
\beq
\bb{ (J\times B)\times B} = \bb{(J\cdot B)B} - B^2\bb{J},
\eeq 
if the currents and fields are perpendicular, ambipolar
diffusion acts like a field-dependent resistivity.  This is the case
for the special vertical field geometry treated by W99 and in \S 4.2
below: all of ambipolar diffusion may be incorporated into an effective
resistivity.  But for the more general field geometries we consider, no
such simplification is possible.  We shall accordingly assume a neutral
number density much in excess of $10^{13}$ cm$^{-3}$, and ignore
ambipolar diffusion.  The form of the induction equation we shall use
henceforth is
\beq\label{finind} {\dd\B\over \dd t} =
\del\btimes\left( \vv \btimes {\B} - \eta\bb{\del\times B} - { c\bb
{(\del\times B)\times B}\over 4\pi  n_e e} \right).  \eeq

\section{Uniform Rotation}

Hall currents introduce novel elements into the MHD of astrophysical
disks, and to develop an intuitive understanding, we consider a very
simple problem: Hall-modified Alfv\'en waves in a uniformly rotating
disk threaded by a vertical magnetic field, $\bb{B} = B\bb{e_z}$.  Here
we shall ignore finite resistivity and vertical structure, work in the
Boussinesq limit, and use standard cylindrical coordinates $(R, \phi,
z)$ with the origin at the disk center.  Finally, we consider plane
wave perturbations that depend only upon $z$.  Linearized quantities
(indicated by $\delta$ notation) are proportional to $\exp(i\omega t -
ikz)$, where $\omega$ is the angular frequency and $k$ is the vertical
wave number.

Under these circumstances, pressure, density, vertical velocity and
vertical magnetic field perturbations all vanish.  We solve for $\delta
v_R$, $\delta v_\phi$, $\delta B_R$ and $\delta B_\phi$.
The linearized radial and azimuthal equations of motion are
\beq\label{aaa}
-i \omega \delta v_R - 2 \Omega \delta v_\phi - {ikB\over 4\pi\rho}  \, \delta B_R
 = 0,
\eeq
\beq\label{bbb}
-i \omega \delta v_\phi  + 2 \Omega \delta v_R - {ikB\over 4\pi\rho}  \, \delta B_\phi
=0.
\eeq
The same components of the linearized induction equation are
\beq
-i\omega \delta B_R + {k^2 Bc\over 4\pi e n_e}\, \delta B_\phi - ikB\, \delta v_R = 0,
\eeq
\beq
-i \omega \delta B_\phi - {k^2 Bc\over 4\pi e n_e}\, \delta B_R - ikB\, \delta v_\phi = 0,
\eeq
Note the symmetry of the problem, in particular the
introduction of a ``Coriolis term'' into the magnetic field equations.
These, of course, are the electromotive forces of the Hall effect.

The combination $k^2 Bc/4\pi e n_e$ will be recognized by students of
plasma physics as the electron whistler frequency.  At characteristic
wavenumbers satisfying $kv_A \sim \Omega$, this frequency is $\sim
\Omega(\Omega/\omega_{c\mu})(n/n_e)$.  Hence, the ionization fraction
below which the whistler branch couples with the disk dynamics is
$(n_e/n) \sim (\Omega/\omega_{c\mu})$.

The ``whistler drift'' of the field lines with respect to the bulk of
the fluid is caused by the motion of the current-bearing electron
charge carriers.  Perturbing the magnetic field induces local currents,
and it is the electron motion in the currents relative to the ions that
causes the field lines to move relative to the ion-neutral fluid.
(Recall that ambipolar diffusion is ignored, so the ions are locked
with the neutrals.)  Note the sense of the field line drift:  if
$\bb{\Omega}$ and $\bb{B}$ are both upward (say), the induced whistler
circular motion of the field lines is right-handed, following the
electrons, {\em opposite} to motion induced by the dynamical Coriolis
force.

How does all this affect the wave response of the fluid?  Consider
first the case of aligned $\bb{B}$ and $\bb{\Omega}$.  The sense of the
field line motion is to slow the dynamical epicycles.  The easiest way
to see this is to note that in the absence of rotation, equations
(\ref{aaa}) and (\ref{bbb}) show that the perturbed velocity vector is
proportional to the perturbed magnetic field vector.  The whistler
waves would impart a circularly polarized component to the velocity
response.  This response is counter to the direction that the dynamical
epicyclic motion is trying to impart, and hence it effectively lowers
the Coriolis force.  As for the magnetic tension, the additional Hall
response is $90^\circ$ out-of-phase with respect to this force, and we
therefore expect these two to add in quadrature.  If, in addition, the
dynamical epicycle is slowed by Hall currents, the return magnetic
tension force should effectively increase.  If the field and angular
velocity are counter aligned, the signs of these effects should
switch.

These expectations are born out in the dispersion relation, which is
most transparently written
\beq\label{disp1}
\omega^2 \pm 2\Omega\omega(1-k^2v^2_H/4\Omega^2) - k^2(v_A^2 + v_H^2) 
=0,
\eeq
where we have introduced the {\em Hall velocity} defined by
\beq 
v_H^2 \equiv {\Omega B c\over 2\pi e n_e}.  
\eeq
Note that although this quantity has dimensions of (velocity)$^2$, {\em
it may be positive or negative.}  It is positive (negative) if
$\bb{\Omega}$ and $\bb{B}$ are oriented in the same (opposite) sense.
We address more general orientations in subsequent sections.

In the dispersion relation (\ref{disp1}), the upper $+$ sign
corresponds to left-handed polarization relative to the magnetic
field, the lower $-$ sign to right-handed polarization.  Whistler waves
exist only for the right-handed branch, and correspond to the limit
$\Omega, n_e \rightarrow 0$.  There are no instabilities in a uniformly
rotating disk, since there is no free energy source.

We postpone until later sections (which include differential rotation)
a full discussion of the solutions to the dispersion relation, but this is a
convenient spot to make contact with waves in a magnetized plasma.
In the limit $\Omega\rightarrow 0$, the dispersion relation becomes
\beq
\omega^2 \pm \omega {k^2 B c\over 4\pi n_e  e} + k^2 v_A^2 = 0,
\eeq
This is precisely the dispersion relation for a uniformly magnetized
plasma with the displacement current and electron inertia ignored
(e.g., Krall \& Trivelpiece 1973).  (The Alfv\'en speed in the above
is dominated by the neutrals, whereas in standard fully ionized plasma
treatments it is determined by the ions, of course.)  Without loss of generality,
we may take $\omega \ge 0$, allowing $\bb{k}$ to determine the propagation
direction.  The positive frequency solutions are
\beq
\omega = \mp {k^2 B c\over8\pi n_e e} +\left( {k^4 B^2 c^2\over 64\pi^2 n_e^2 e^2}
+ k^2 v_A^2 \right)^{1/2}.
\eeq
For small wavenumbers (low frequencies) both of the above solutions
reduce to Alfv\'en waves.  At large wavenumbers, right-handed waves
($+$ sign) go over to the high frequency whistler wave branch, whereas
large $k$ left-handed waves ($-$ sign) are cut off at a frequency given
by
\beq
\omega\> {\rm (cut\>off)} = {eB\over \mu c} ({n_e\over n}) = \omega_{c\mu}
({n_e\over n}),
\eeq
analogous to the ion-cyclotron cut-off in a fully ionized plasma.
The characteristic ionization fraction at which the cut-off frequency
drops below the disk rotation frequency is once again $(n_e/n) \sim
\Omega/\omega_{c\mu}$.

%To summarize: for the simple field geometry under consideration here,
%aligned magnetic and angular velocity vectors result in Hall currents
%that retard the dynamical epicyclic motion, but counter-aligned vectors
%enhance this motion.  In the next section, we investigate how these
%effects play out in a differentially rotating disk.

\section {Differential Rotation}

\subsection {Preliminaries.}

We consider next the local stability of a differentially rotating disk
threaded by a weak vertical field.   We assume that finite resistivity
and Hall currents are both present; we neglect ambipolar diffusion.  As
in the previous section, we restrict ourselves to plane wave
disturbances of the form $\exp( i k z - i \omega t)$.  In the
Boussinesq limit, this corresponds to fluid displacements in the plane
of the disk, so vertical structure is unimportant.

This problem has been considered by W99, but it bears re-examination
for several reasons.  It is an important baseline for understanding the
effects of Hall currents in protostellar disks, and is worth
understanding from more than one perspective.  Our approach emphasizes
dynamical couplings rather than conductivity tensor formalism, and is
physically quite distinct.  It is also somewhat simpler.  Finally, we
shall examine a qualitatively new feature of the Hall effect that has
not been discussed before: HEMFs can destabilize {\em outwardly}
increasing differential rotation.  In the process we show that the
maximum growth rate of the instability is always the local Oort A value
of the rotation (in magnitude), whether the angular velocity is
increasing or decreasing outward.

\subsection {Axial Fields and Wavenumbers}
\subsubsection{Stability}

The key issue of the analysis is the local stability criterion of a disk
with an axial field.  
Whereas the dispersion relation is somewhat complex, the limit $\omega
\rightarrow 0$ is simple, and this is relevant for understanding local
stability.  (In principle, local overstability may occur, but for
self-consistency we wish to focus on nonpropagating, evanescent, local
modes.  For these, the transition between stability and instability
must proceed through the point $\omega = 0$.)

Including the effects of differential rotation and finite resistivity
along with the Hall terms, we find that the linearized, $\omega = 0$,
radial and azimuthal equations of motion are now

\beq\label{ad}
 - 2 \Omega \delta v_\phi - {ikB\over 4\pi\rho}  \, \delta B_R
 = 0,
\eeq
\beq\label{ac}
{\kappa^2\over 
2 \Omega} \delta v_R - {ikB\over 4\pi\rho}  \, \delta B_\phi =0,
\eeq
where $\kappa$ is the epicyclic frequency, defined by
\beq
\kappa^2 = 4\Omega^2 + {d\Omega^2\over d\ln R}.
\eeq
The same components of the linearized induction equation are
\beq\label{ab}
 k^2\eta \delta B_R + {k^2 Bc\over 4\pi e n_e}\, \delta B_\phi - ikB\, \delta v_R = 0,
\eeq
\beq\label{aa}
k^2 \eta \delta B_\phi - \left( {k^2 Bc\over 4\pi e n_e} +
{d\Omega\over d\ln R}\right) \delta B_R - ikB\, \delta v_\phi = 0,
\eeq
Notice that in equation (\ref{aa}), the whistler frequency
Hall term introduces a coupling
that is equivalent to changing the shear rate $d\Omega/dR$.  It arises
becauses the $\phi$ component of the perturbed electron velocity
($\delta \bb{v_e}$) differs from the dominant neutral velocity by a
term involving $\delta J_\phi$.  This produces a radial field
component, $\delta B_R$.  The effect is present whether or not there is
differential rotation, but it is most striking when such motion is
present, because it couples exactly like $d\Omega/dR$.  In effect,
it is the magnetic ``epicyclic'' term, combining a Coriolis-like
coupling with a background angular velocity gradient.

If we ignore resistivity for the moment, our problem decouples into two
very simple subproblems, one with $\delta v_\phi$ coupled only to
$\delta B_R$, and one with $\delta v_R$ coupled only to $\delta
B_\phi$.  The torque must vanish, so equations (\ref{ac}) and
(\ref{ab}) combine to give
\beq
\left( k^2 v_A^2 + {\kappa^2 k^2 v_H^2\over4\Omega^2}\right) \delta B_\phi = 0,
\eeq
or the tidal and excess centrifugal forces must vanish, in which case
equations (\ref{ad}) and (\ref{aa}) yield
\beq
\left( {d\Omega^2\over d\ln R} + k^2( v_A^2 + v_H^2) \right)\delta B_R = 0 .
\eeq
If we are not to have a trivial solution, then
\beq
\left( v_A^2 + {\kappa^2  v_H^2\over4\Omega^2}\right) 
\left( {d\Omega^2\over d\ln R} + k^2( v_A^2 + v_H^2) \right) = 0.
\eeq
In the standard MRI, the torque is always purely Alfv\'enic, and
therefore it always has the same sign: a restoring negative torque for
a positive angular displacement.  This means that instability depends
only upon the direction of the excess centrifugal force.  The Hall
effect renders the problem more interesting, allowing an interplay
between the torque and centrifugal force.  An {\em inwardly} directed
centrifugal force can destabilize if accompanied by a positively
directed torque.

The inclusion of resistivity couples $\delta B_R$ and $\delta B_\phi$,
but gives only a slight modification to the above result:
\beq\label{stab1}
\left(  v_A^2 + {\kappa^2 v_H^2\over4\Omega^2}\right)
\left( {d\Omega^2\over d\ln R} + k^2( v_A^2 + v_H^2) \right)  + \kappa^2 \eta^2
k^2 = 0 .
\eeq
This is the desired critical stability condition.  To determine the sign for
instability, one may take the limit
$v_H^2, \eta \rightarrow 0$, which returns us to the simple MRI.
It then easily follows that the left hand side
should be negative for instability, that is  
\beq\label{inequ}
k^2 v_A^2 \left[ (1+x)( 1 + {\kappa^2 x\over 4\Omega^2}) + {\kappa^2\eta^2\over
v_A^4} \right] < - \left[ 1 + {\kappa^2 x \over 4\Omega^2}\right]
{d\Omega^2\over d\ln R}
\eeq
where
\beq
x \equiv {v_H^2\over v_A^2}
\eeq
is the dimensionless {\em Hall parameter.}   The terms in square brackets incorporate
the additional physics of HEMFs and resistivity; they become
unity for the standard $x\rightarrow 0$ MRI (Balbus \& Hawley 1991).
The dimensionless resistivity parameter may be expressed in terms of
the magnetic Reynolds number:
\beq
{\kappa\eta\over v_A^2} = \big({\kappa\over \Omega}\big) \big({c_S\over v_A}\big)
{Re_M}^{-1}.  
\eeq

The behavior of the polynomial
\beq
D(x) = (1+x)( 1 + {\kappa^2 x\over 4\Omega^2}) + {\kappa^2\eta^2\over
v_A^4}
\eeq
determines the nature of the instability.  We may write the instability criterion
\beq\label{fifty}
{\rm sgn}(D) k^2 v_A^2  < - { (1 + \kappa^2 x/4\Omega^2) {d\Omega^2/d\ln R} \over
|D|}.
\eeq
It is not difficult to show that if
\beq\label{window}
{Re_M}^{-1} < {1\over4} 
\left(\Omega^2\over \kappa^2\right)  \left(v_A\over c_S\right)
\left| d\ln\Omega^2\over d\ln R\right|
\eeq
there will be a finite range of $x$ for which $D(x)<0$.   In this
window, all wavenumbers, no matter how large, are unstable, even in the
presence of resistivity: the right hand side of the inequality
(\ref{inequ}) is positive, while the left hand side is negative.
Clearly, the zeroes of $D$ are critical for understanding the behavior
of disks in the Hall regime.

For a Keplerian disk, the window occurs when $x$ falls approximately
between $-1$ and $-4$ for large $Re_M$.  Roughly speaking, there is a
match between the response frequency of the magnetic tension and the
``epicyclic'' frequency of the field line drift through the fluid
caused by HEMFs.  These tend to cancel each other's dynamical
response.  Qualitatively, it is as though the restoring tension were
not present as a radial force, leaving the destabilizing dynamical tide
to do its work.  Within the window, going to larger wavenumbers does
not change the sign of the radial forces on a fluid element.  The
azimuthal torque, however, is still dominated by the Alfv\'en term, and
this allows for angular momentum transfer.

\begin{figure}
\plotone{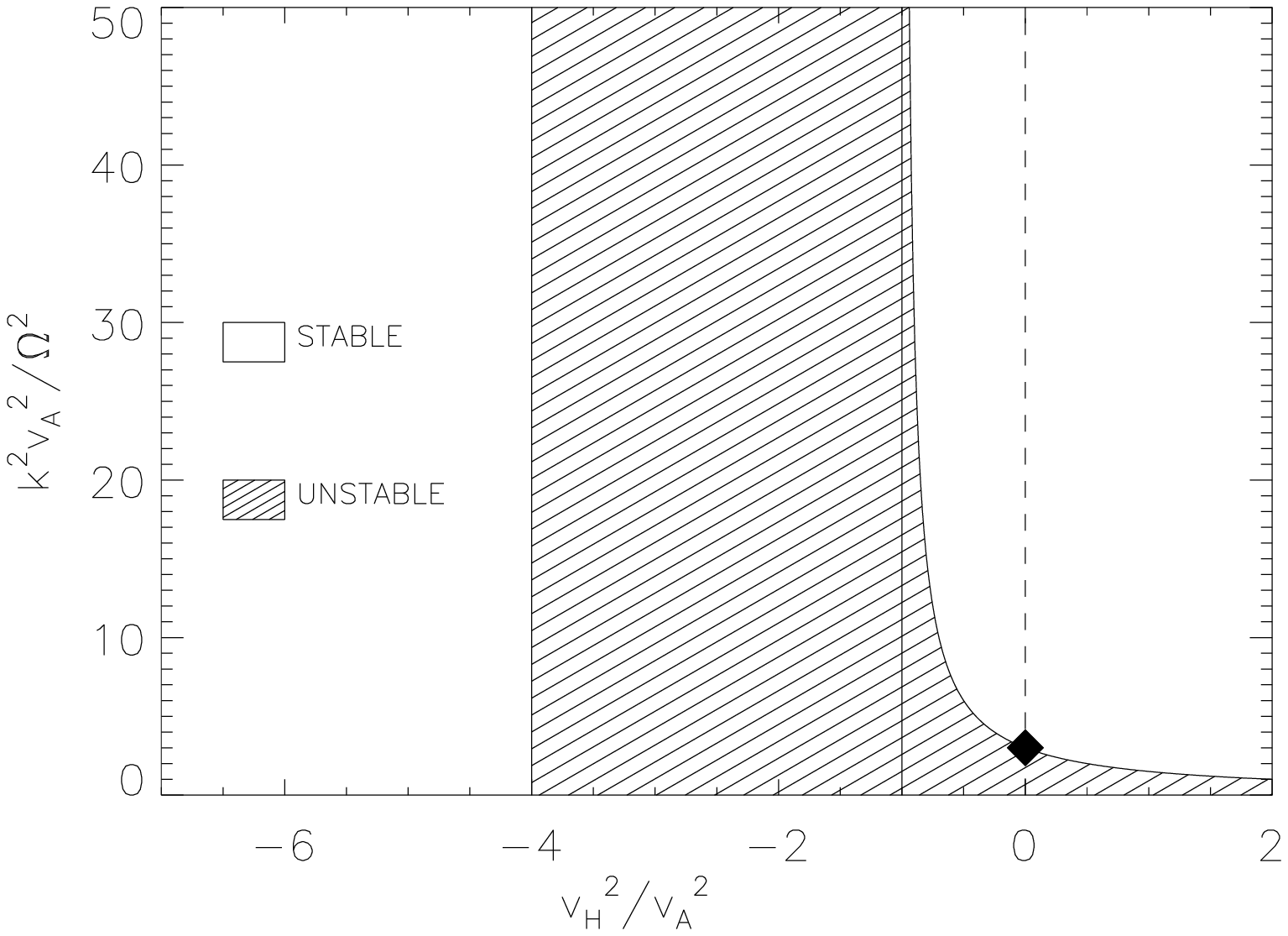}
\caption[] {Stable and unstable regions in the $(kv_A/\Omega)^2,\>
x\equiv (v_H/v_A)^2$ parameter space for the case of zero resistivity.
The black diamond corresponds to the
location of the critical axial wavenumber for the MRI.  Positive values
of $v_H^2$ always stabilize because the right-handed polarization of whistler waves
decreases the effective magnitude of the destabilizing shear.  Negative
values at first destabilize by effectively increasing the shear rate,
but ultimately stabilize in the form of whistler waves at large magnitudes
of the Hall parameter.  Notice that for $-4\le x\le -1$, all wavenumbers are
unstable; magnetic tension is eliminated as a stabilizing agent.}
\label{wfig1}
\end{figure}

In figure (\ref{wfig1}), we show a stability plot for an $\eta =0$
Keplerian disk in the $x, (kv_A/\Omega)^2$ plane.  The shaded regions
correspond to instability.  The instability window (i.e., the region in
which all wavenumbers are unstable) for $-4\le x \le -1$ is evident,
together with an abrupt transition for $x< -4$.  For $0> x> -1$ the
Hall currents destabilize by allowing for a larger range of
destabilizing wavenumbers than would be present in the
magnetorotational instability.  For $x>0$, Hall currents are a
stabilizing influence.

\begin{figure}
\plotone{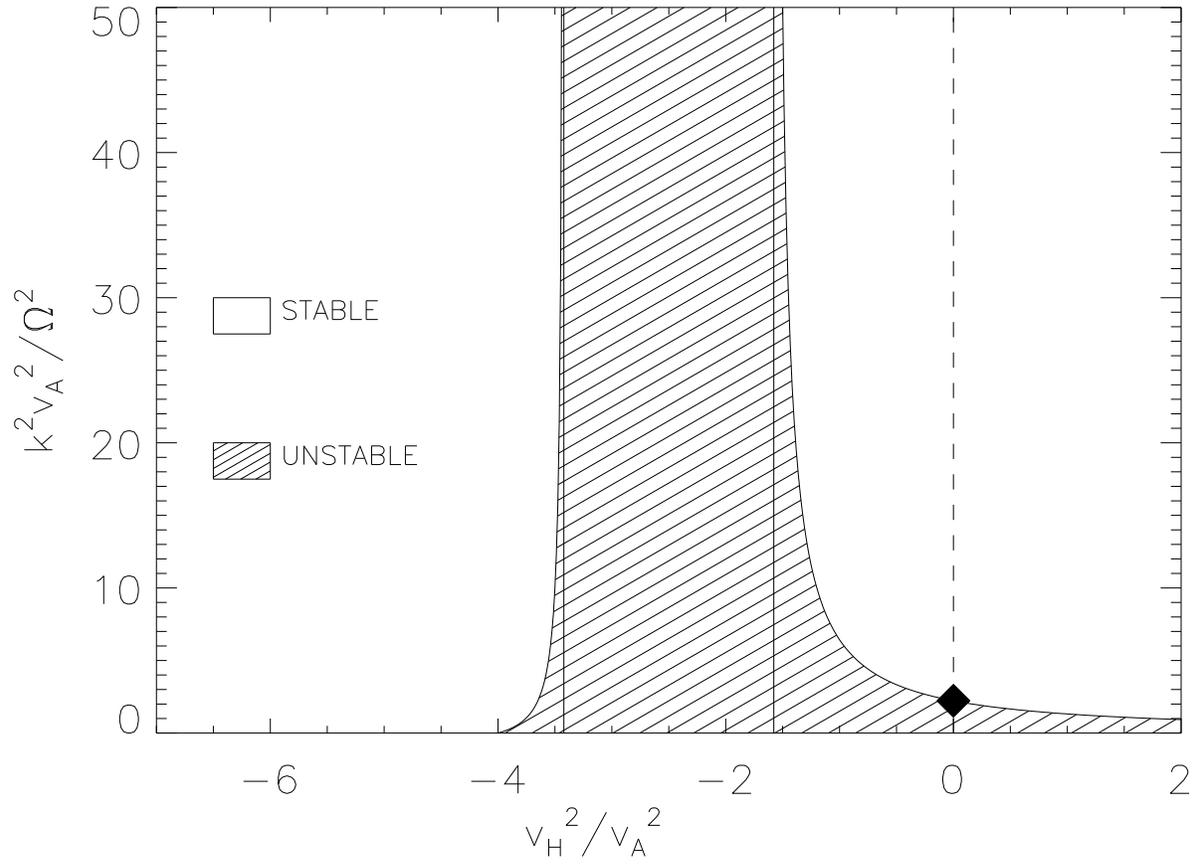}
\caption[] {As in fig.~(1), but with finite resistivity,
$\kappa^2\eta^2/v_A^4 = 0.35$.
The window of instability at all wave numbers is still present even
with resistance, but it is more narrow.}
\label{wfig2}
\end{figure}

Figure (\ref{wfig2}) illustrates the modified instability range for large
$Re_M$.  The window is narrowed as shown.

\begin{figure}
\plotone{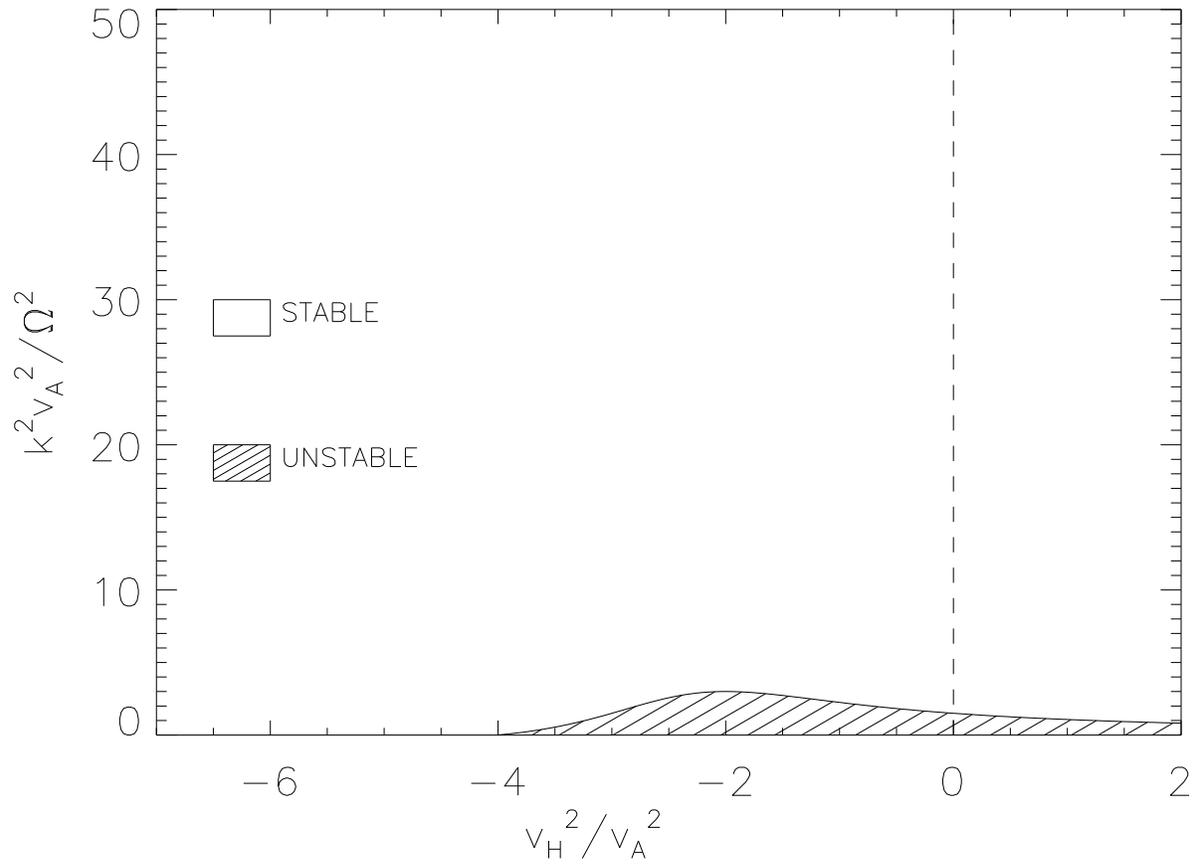}
\caption[] {As in fig.~(2), but with larger resistivity,
$\kappa^2\eta^2/v_A^4 = 1$.  The window of instability at all wave
numbers has vanished, and only the largest wavelengths remain formally
unstable.}
\label{wfig3}
\end{figure}

Figure (\ref{wfig3}) illustrates the effect of increased resistivity on
stability.  The window has now disappeared, and as $Re_M$ diminishes,
an ever decreasing portion of wavenumber space is unstable.  This
region corresponds to very large wavelengths, which may exceed the
global disk scales.  This corresponds to resistive stabilization.  In
general, when resistivity becomes dominant, the instability criterion
(\ref{fifty}) leads to $kv_A < O(Re_M^{-1})$.  But the the right hand side
of the inequality is a function of $x$, and at its maximum we find
$kv_A < O(Re_M{-1/2})$.  This is considerably easier to satisfy, though
it corresponds to tuning the Hall parameter.  This effect was also
noted in W99.

\begin{figure}
\plotone{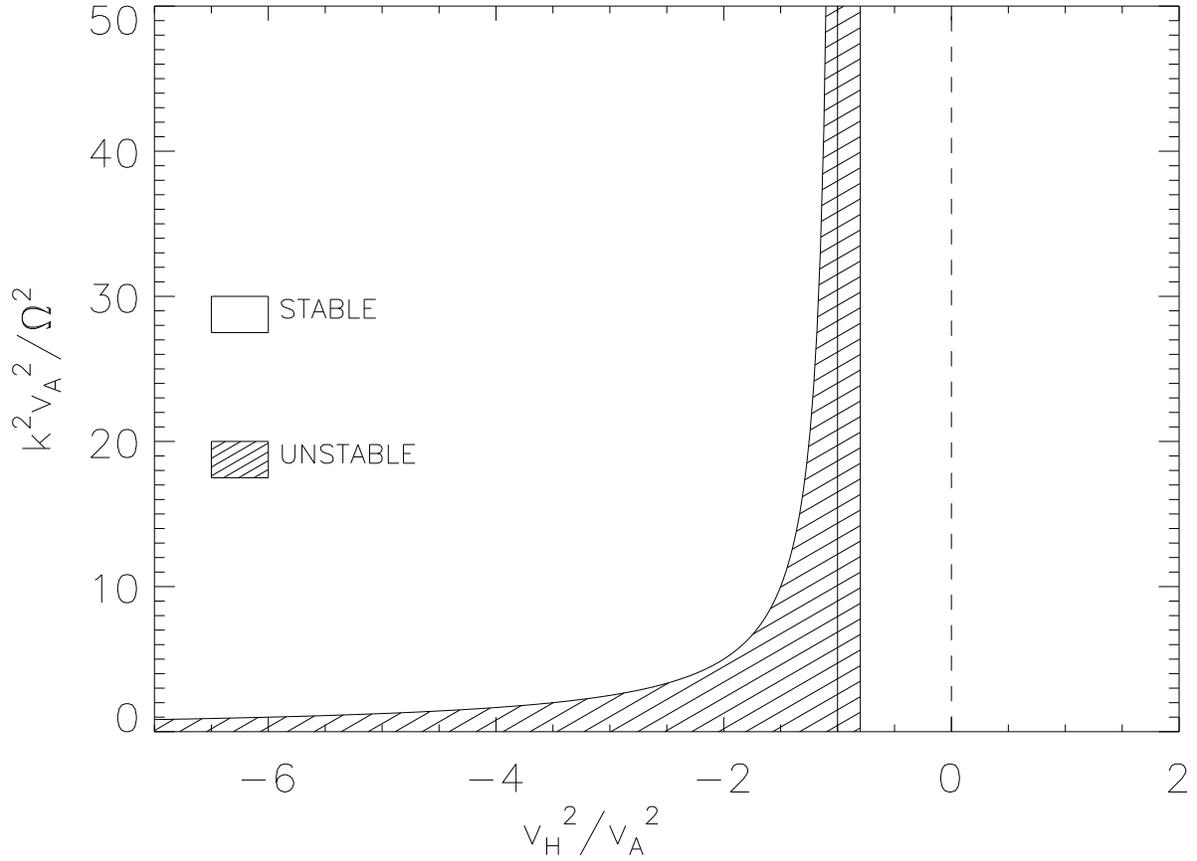}
\caption[] {Instability diagram of an $\eta = 0$,
$\kappa^2 = 5\Omega^2$ disk
profile, corresponding to an outwardly increasing rotation profile.
No instability is present in such a disk without the Hall EMF.
With it, the disk is unstable, and shares what seems to be
the universal maximum growth rate, $0.5\, d\Omega/d\ln R$.}
\label{wfig4}
\end{figure}

Figure (\ref{wfig4}) is presented to illustrate a point of principle.
It is the stability diagram of an $\eta = 0 $, $\kappa^2 = 5\Omega^2$
(i.e., $d\Omega^2/dR > 0$) disk.  There is an instability window,
analogous to its Keplerian counterpart, for $-4\Omega^2/\kappa^2> x >
-1$, abrupt stability for $x > -4\Omega^2/\kappa^2$, and instability
for an ever diminishing wavenumber domain as $x$ falls below $-1$.  In
the presence of sufficiently large Hall currents, with counteraligned
axial angular velocity and magnetic field vectors, the instability
criterion is $d\Omega^2/dr \not= 0$.  Any differential rotation law is
potentially unstable.

\subsubsection{Dispersion Relation}

We next obtain the full dispersion relation.
Assume that the time dependence of the perturbations
takes the form $\exp(\sigma t)$.  (This form keeps the 
coefficients of the dispersion relation real.)
The linearized radial and azimuthal equations of motion are now
\beq
\sigma \delta v_R - 2 \Omega \delta v_\phi - {ikB\over 4\pi\rho}  \, \delta B_R
 = 0,
\eeq
\beq
 \sigma \delta v_\phi  +{\kappa^2\over 
2 \Omega} \delta v_R - {ikB\over 4\pi\rho}  \, \delta B_\phi =0.
\eeq
The same components of the linearized induction equation are
\beq
(\sigma +  k^2\eta) \delta B_R + {k^2 Bc\over 4\pi e n_e}\, \delta B_\phi - ikB\, \delta v_R = 0,
\eeq
\beq
(\sigma +  k^2 \eta)
\delta B_\phi - \left( {k^2 Bc\over 4\pi e n_e} +
{d\Omega\over d\ln R}\right) \delta B_R - ikB\, \delta v_\phi = 0,
\eeq
The resulting dispersion relation is 
\beq\label{basdis}
0=\sigma^4 + 2 \eta k^2\sigma^3 + {\cal C}_2 \sigma^2 +
2 \eta k^2 (\kappa^2 + k^2 v_A^2) \sigma + {\cal C}_0
\eeq
where the constants ${\cal C}_2$ and ${\cal C}_0$ are given by 
\beq
{\cal C}_2 = 
\left[
\kappa^2 + 2 k^2 v_A^2 + \eta^2 k^4 + {k^2 v_H^2\over 4\Omega^2}
({d\Omega^2 \over d\ln R} + k^2 v_H^2 )\right],
\eeq
and%  ${\cal C}_0$ is given by the left side of equation (\ref{stab1}).
\beq
{\cal C}_0 = k^2 \left[ v_A^2 + {\kappa^2v_H^2\over 4\Omega^2}\right]
\left[ {d\Omega^2\over d\ln R} + k^2( v_A^2 + v_H^2) \right]
+ \kappa^2 \eta^2 k^4.
\eeq

Balbus \& Hawley (1992a) conjectured that the maximum growth rate of any
instability feeding off the differential rotation in a disk is given by
the local Oort A value, $\sigma_A\equiv(1/2)|d \Omega/ d\ln R|$.  All
magnetic field configurations of the magnetorotational instability have
this value for their most rapid growth rate.  However, the A value
conjecture went beyond Alfv\'en tension as the destabilizing agent, it
suggested that whatever the proximate cause, linear perturbations can
grow no faster than $\sigma_A$.  The reasons are rooted in the dynamics
of the differential rotation process itself, not in magnetism {\em per
se.}  Now, it is far from obvious that in the absence of resistivity, the
rather unwieldy dispersion relation (\ref{basdis}) has precisely this
value as its maximum growth rate.  But it does, as we now demonstrate.

To begin, write equation (\ref{basdis}) in dimensionless form, with all
rates normalized to $\Omega$.  We define
\beq
s = \sigma/\Omega, \quad{\tilde\kappa} = \kappa/\Omega,
\quad X=(kv_A/\Omega)^2, \quad Y= (kv_H/\Omega)^2.
\eeq
Note that
\beq\label{help}
{\tilde\kappa}^2 = 4 + {d\ln \Omega^2\over d\ln R}.
\eeq
The $\eta = 0$ dispersion relation is
\beq\label{dissimp}
s^4 + \left[{\tilde\kappa}^2 + 2X +{Y\over 4} \left( {d\ln\Omega^2\over d\ln R}+Y
\right)\right]s^2 + \left( X + {Y{\tilde\kappa}^2\over 4}\right)\left(
{d\ln\Omega^2\over d\ln R} +X+Y\right) = 0.
\eeq
At the maximum growth rate $s=s_m$, partial differentiation of the above with respect
to $X$ and $Y$ gives the two equations
\beq\label{parX}
X + {Y\over 8}({\tilde\kappa}^2 + 4)  = -s_m^2 - {1\over 2} {d\ln\Omega^2\over d\ln
R}
\eeq
\beq\label{parY}
{X\over 4}({\tilde\kappa}^2 + 4) + {Y\over2} ({\tilde\kappa}^2 + s_m^2) =
-{1\over4}{d\ln\Omega^2\over d\ln R}({\tilde\kappa}^2 + s_m^2).
\eeq
Using the identity (\ref{help}) and
eliminating $Y$ between equations (\ref{parX}) and (\ref{parY}),
leads after simplification to a remarkable result,
\beq
\left(X + s_m^2 + {\tilde\kappa}^2\right)\left[s_m^2 -
{1\over16}\left(d\ln\Omega^2\over d\ln R\right)^2 \right] = 0.
\eeq
Since the first factor must be positive definite, we conclude that
\beq
s_m^2 = {1\over 16} \left( d\ln\Omega^2\over d \ln R\right)^2.
\eeq
This is equivalent to 
\beq
s_m = {1\over 2} \left| d\ln \Omega\over d\ln R \right|,
\eeq
i.e., the Oort A value!  

We must now verify that this value of $s_m^2$ is a solution of the
dispersion relation (\ref{dissimp}) for $X$ and $Y$ satisfying
the system of equations (\ref{parX}) and (\ref{parY}).   It is easily
shown that with the Oort A value chosen for $s_m$, the
determinant of this system vanishes.   We thus use only one of these
equations, which we take to be (\ref{parX}).    First we substitute
for $s_m^2$, and solve for $X$ in terms of $Y$.
This leads to
\beq
X= - {1\over 2} {d\ln\Omega^2\over d\ln R} - {Y\over8} ({\tilde\kappa}^2 +4)
-{1\over16} \left(d\ln\Omega^2\over d\ln R\right)^2.
\eeq
Note that in the $Y=0$ limit of this equation, $X\ge 0$ is possible
only for decreasing outward angular velocity profiles.  The
magnetorotational instability requires $d\Omega^2/dR <0$.  The addition
of the Hall effect allows for the possibility that $Y<0$, and thus for
a perfectly well-defined positive $X$ parameter, even if $d\Omega^2/dR
> 0$.  That both senses of angular velocity gradient may be treated
on the same footing, accords with the well-known result that the state
of minimum energy for a disk of fixed angular momentum is solid body
rotation (Lynden-Bell \& Pringle 1974).  In principle {\em any} angular
velocity gradient could be unstable, since lower energy states
consistent with angular momentum conservation exist.  A dynamical path
to these low energy states is required however, and that is what HEMFs
can provide.

Replacing $X$ by the above
expression wherever it occurs in equation (\ref{dissimp})
yields, after some simplification, 
\beq
s^4 + c_2 s^2 + c_0 = 0,
\eeq
with 
\beq
 c_2 = {1\over4} (Y-4)^2 - {1\over 8} \left( d\ln\Omega^2\over d\ln R\right)^2,
\eeq
and
\beq
c_0 = {1\over 64}\left( d\ln\Omega^2\over d\ln R\right)^2
\left[ {1\over4}\left( d\ln\Omega^2\over d\ln R\right)^2
-(Y-4)^2 \right]
\eeq
The standard solution is
\beq
s^2 = {1\over8} 
\left[ {1\over2} \left( d\ln\Omega^2\over d\ln R\right)^2-(Y-4)^2
\pm (Y-4)^2 \right]
\eeq
Choosing the $+$ sign returns the Oort A solution, exactly as desired
and consistent with the Balbus \& Hawley (1992a) conjecture.
(W99 obtained the Keplerian upper limit by numerical solution of the 
dispersion relation, and noted that it was the same as the standard
magnetorotational instability.) 

\begin{figure}
\plotone{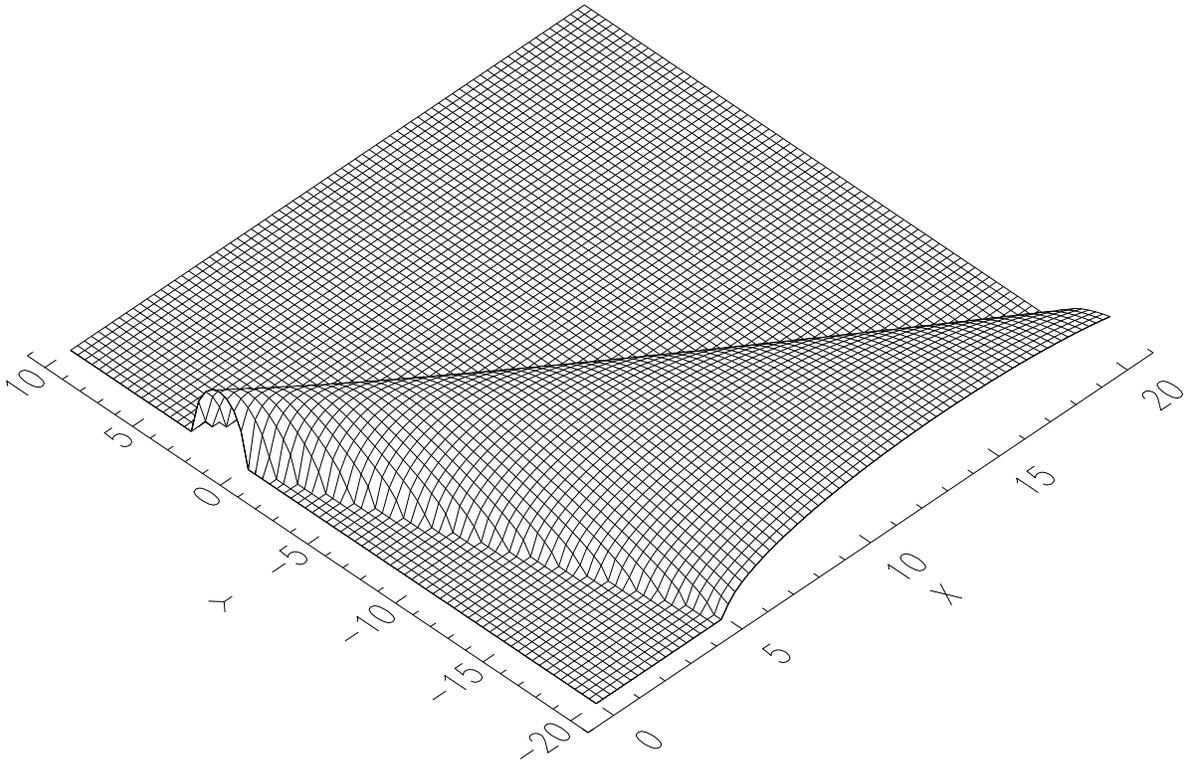}
\caption[] {Growth rate in the $X, Y$ plane for an $\eta = 0$ Keplerian
disk.  (See text for definitions.)  Only regions of instability are
shown, with the height proportional to the growth rate.  The vertical
axis has been suppressed, but the maximum growth rate of the ``ridge''
is 0.75 $\Omega$.  $Y<0$ corresponds to counter-aligned $\bb{\Omega}$
and $\bb{B}$.}
\label{wfig5}
\end{figure}

\begin{figure}
\plotone{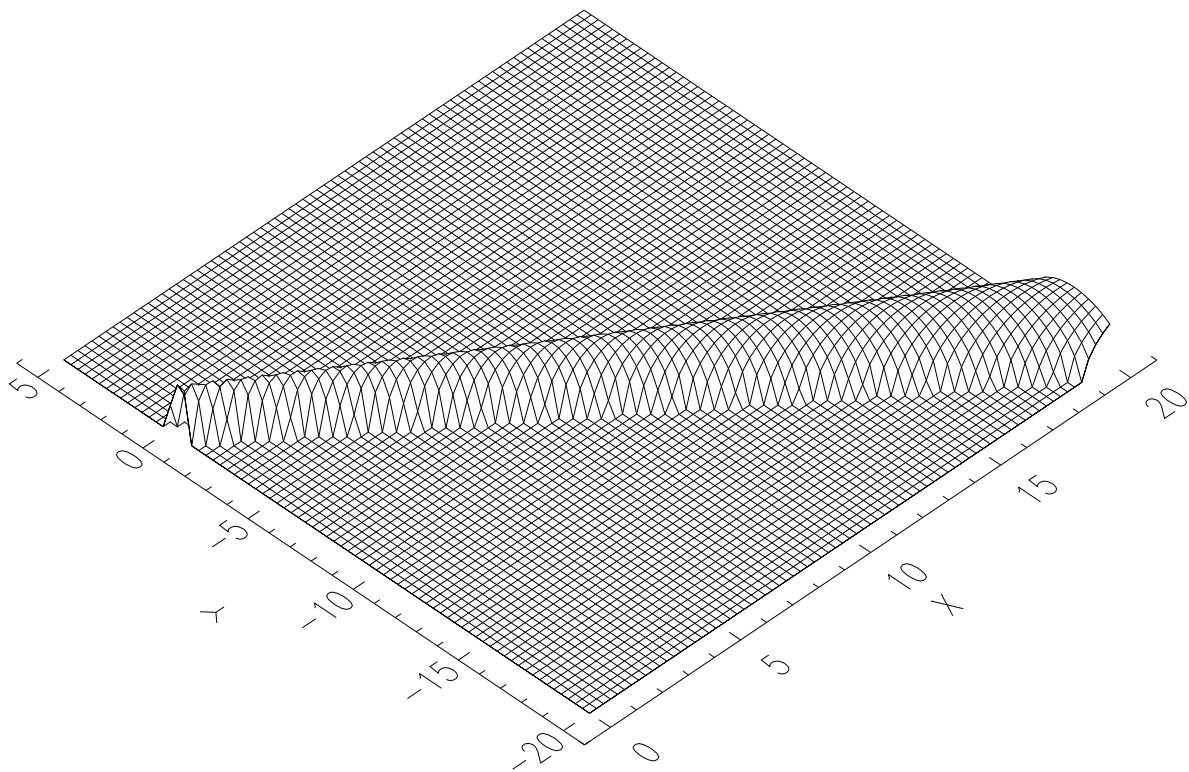}
\caption[] {As in fig.~(5), but with $\kappa^2 = 5\Omega^2$, a disk with an
outwardly increasing velocity profile.  Only $Y<0$ regions can be unstable.}
\label{wfig6}
\end{figure}

Graphical solutions in the $XY$ plane are shown in figures (\ref{wfig5}) and
(\ref{wfig6}).

\subsection{General Axisymmetric Disturbances}

We now consider the axisymmetric behavior of the instability with more
general field geometries and wavenumbers.   We shall ignore buoyancy,
so our analysis holds either for a strictly polytropic disk, or locally
at the midplane for any constituative relation.  The perturbation
wavevector is
\beq
\bb{k} = k_R \bb{ {e_R}} + k_Z \bb{ {e_Z} },
\eeq
and disturbances have space-time dependence $\exp(i\bb{k\cdot r}+\sigma t)$.
The linearized equations are now that of mass conservation (Boussinesq limit):
\beq
k_R \delta v_R + k_Z \delta v_Z = 0 ,
\eeq
and the equations of motion,
\beq
\sigma \delta v_R - 2 \Omega \delta v_\phi - {ik_Z B_Z\over 4\pi\rho}  \, \delta B_R
+ { ik_R\over 4\pi\rho} (B_\phi\, \delta B_\phi +B_Z\delta B_Z)+
ik_R{\delta P\over \rho}  = 0,
\eeq
\beq
\sigma \delta v_\phi  +{\kappa^2\over
2 \Omega} \delta v_R - {i\bb{(k\cdot B)}\over 4\pi\rho}  \, \delta B_\phi =0,
\eeq
\beq
\sigma \delta v_Z  - {i k_R B_R \over 4\pi \rho} \, \delta B_Z
+ {ik_Z\over 4\pi \rho} (B_\phi\, \delta B_\phi +B_R\delta B_R)
+ {i k_Z \delta P\over \rho} = 0.    
\eeq
If one considers a polytropic disk, then $\delta P/\rho$ may be
replaced directly by a pure enthalpy perturbation, $\delta Q$,  say.
Otherwise it is to be taken at face value, a pressure perturbation
divided by the density $\rho$.  Ultimately, either choice leads to the
same result, because the term serves only as a place holder.

The linearized induction equations are
\beq
(\sigma +  k^2\eta) \delta B_R + {c \bb{(k\cdot B)} k_Z \over 4\pi e n_e}\, \delta B_\phi - i
\bb{(k\cdot B)} \delta v_R = 0, 
\eeq
\beq
(\sigma + k^2\eta) \delta B_\phi - \left( {d\Omega\over d\ln R} + {c \bb{(k\cdot B)} k_Z
\over 4\pi e n_e}\right)\, \delta B_R + {c \bb{(k\cdot B)} k_R \over 4\pi e n_e}\, \delta
B_Z - i \bb{(k\cdot B)} \delta v_\phi = 0, 
\eeq
\beq
(\sigma +  k^2\eta) \delta B_Z - {c \bb{(k\cdot B)} k_R \over 4\pi e n_e}\, \delta B_\phi
- i \bb{(k\cdot B)} \delta v_Z = 0.  
\eeq
The dispersion relation that emerges after a straightforward but lengthy effort is
\beq\label{gendisp}
\sigma^4 + 2 \eta k^2 \sigma^3 + {\cal C}_2 \sigma^2 + 2\eta k^2 \left( {k_Z^2\over k^2}
\kappa^2 + \bb{(k\cdot v_A)}^2 \right) \sigma + {\cal C}_0 =0,
\eeq
with
\beq\label{gendispp}
{\cal C}_2 = {k_Z^2\over k^2} \kappa^2 + 2 \bb{(k\cdot v_A)}^2 + \eta^2  k^4
+ {c \bb{(k\cdot B)} k_Z\over 8\pi \Omega e n_e} \left( {d\Omega^2\over d\ln R}
+ {c k_Z\Omega \bb{ (k\cdot B)}\over 2 \pi e n_e} {k^2\over k_z^2} \right),
\eeq
and
\beq \label{gendisppp}
{\cal C}_0 =
\eta^2 k_Z^2 k^2 \kappa^2 + \left( (\bb{k\cdot v_A})^2 +
{c k_Z\Omega (\bb{ k\cdot B})\over 2 \pi e n_e}
+ {k_Z^2\over k^2}{ d\Omega^2\over d\ln R} \right)
\left( (\bb{k\cdot v_A})^2 + {\kappa^2\over4\Omega^2} 
{c k_Z\Omega (\bb{ k\cdot B})\over 2 \pi e n_e}
\right)
\label{inst1}
\eeq 
A sufficient criterion for instability is clearly
\beq\label{genins}
{\cal C}_0 <0.
\label{inst2}
\eeq
This would ensure that the sign of the quartic on the left side of
equation (\ref{gendisp}) changes as $\sigma$ passes from small to large
positive values.  Hence there would have to be a positive root
somewhere.  The necessity of (\ref{genins}) as an instability criterion
is more difficult to prove, but is very likely to be true.  It is
straightforwardly proven in the limit of vanishing resistivity that
${\cal C_0}\ge 0$ leads to stability (since the analysis reduces to the
roots of a quadratic), and the presence or resistivity will almost
certainly cause further stabilization.  Numerical analyses are all
consistent with (\ref{genins}) as a necessary and sufficient criterion
for instability.

The most important difference between strictly axial geometry and the
present calculation lies with the form of the coupling of the Hall term
and wavenumber.  Whereas the axial result makes it appear that the Hall
term stabilizes or destabilizes according to whether $\bb{\Omega\cdot
B}$ is positive or negative, equations
(\ref{gendisp})--(\ref{gendisppp}) show that the coupling is actually
$(\bb{ k\cdot\Omega})(\bb{k\cdot B})$.  The significance of this form
of the coupling is that under conditions of marginal stability, if
there is any finite radial field component, there will {\em always} be
wavenumbers which make this term negative (destabilizing).  This is of
particular importance for the nonlinear development of the instability,
where power is cascaded throughout the wavenumber spectrum.

Consider the case of astrophysical interest, $d\Omega^2/dR < 0$.  If the final
factor of equation (\ref{gendisppp}) (the torque term) is negative, the
preceding factor (the centrifugal term) will also be negative, and obviously
no instability will be possible.  Let us assume, therefore, that the torque is
positive.  Then, a necessary condition for instability is
\beq\label{necc}
(\bb{ k\cdot v_A})^2 + {c k_Z\Omega (\bb{ k\cdot B})\over 2 \pi e n_e}
< - {k_Z^2\over k^2} {d\Omega^2\over d\ln R},
\eeq
which differs from the standard MRI only by an
additive Hall term on the left hand side.  (This is a necessary condition
because it omits resistivity.)
The effect of this is shown 
schematically in figure (7).
%and representative solutions are shown
%in figures (X)-(X), with positive, negative, and vanishing Hall term.
%The role of the dimensionless parameter $(v_H/v_A)^2$ is now played by
%\beq
%{c k_Z \Omega \bb{(k\cdot B)}\over 2\pi e  n_e (\bb{ k\cdot v_A})^ 2 }
%= {2\rho c k_Z\Omega\over e n_e \bb{(k\cdot B)}}.
%\eeq
We have introduced the dimensionless Hall parameter,
\beq
Ha\equiv {c k_Z (\bb{k\cdot B})\over 2\pi e  n_e \Omega},
\eeq
the characteristic whistler frequency normalized to $\Omega$.

Let us finally note that HEMFs will tend to stabilize geometries with
vanishing axial field.  In the absence of the Hall term, the standard
$B_Z=0$ MRI, has its maximum growth rate for disturbances with
$k_Z\rightarrow \infty$.  But with Hall EMFS included, arbitrarily
large axial wavenumbers will not allow ${\cal C}_0 < 0$.  Instead,
these modes become stable waves.

\subsection {Nonaxisymmetric Disturbances}

\subsubsection {Shearing Coordinates}

One of the most important reasons for obtaining this general
axisymmetric formula (\ref{gendisp}) is that it provides a basis for
understanding the behavior of local {\em nonaxisymmetric} disturbances,
to which we now turn.

We shall work in the local shearing coordinate system introduced by 
Goldreich \& Lynden-Bell (1965) in their pioneering study of local
gravitational disk instabilities, and used
in a magnetic context by Balbus \& Hawley (1992b) and Terquem \& Papaloizou
(1996).  The shearing coordinates are defined by
\beq
R' = R, \qquad \phi' = \phi - \Omega(R) t, \qquad z' = z,
\eeq
which leads to
\beq
{\dd\ \over\dd R} = {\dd\ \over\dd R'} - t{d\Omega\over dR} {\dd\ \over\dd 
\phi'}, \qquad {\dd\ \over\dd\phi} = {\dd\ \over \dd\phi '}, \qquad 
{\dd\ \over\dd Z} = {\dd\ \over\dd Z'}.
\eeq
The partial time derivative must be taken holding shear coordinates constant,
which is just the Lagrangian time derivative for the undisturbed disk flow:
\beq
{d\ \over dt} = {\dd\ \over \dd t'} = {\dd\ \over\dd t} + \Omega(R){\dd\ \over
\dd\phi}.
\eeq
Local disturbances depend on these sheared coordinates as $\exp[i(k'_R R' +m\phi'+
k_z z')]$, where the wavevector $\bb{k'}$ is a constant.  This means that
we may effect the transformation from Eulerian to (unperturbed) Lagrangian
coordinates simply by replacing the static Eulerian wavevector $k_R$ with
\beq
k_R(t) = k'_R - mt {d\Omega\over dR},
\eeq
and calculating the time evolution with the Lagrangian derivative. 
Throughout this section, $\bb{k}$ shall denote the shearing wavevector
$(k_R(t), m/R, k_Z)$. 

In the presence of shear, the toroidal magnetic field grows linearly with time:
\beq
B_\phi(t) = B_\phi(0) + t\, B_R\, {d\Omega\over dR}.
\eeq
Note, however, that $\bb{k\cdot B}$ is time-independent:
\beq
\bb{k\cdot B} = k'_R B_R + {mB_\phi(0)\over R} + k_Z B_Z.
\eeq
This has important consequences for nonaxisymmetric behavior. 

\subsubsection{Linear Equations}

The linearized dynamical equations are
\beq
k_R \delta v_R + {m\over R} \delta v_\phi + k_Z \delta v_Z = 0,
\eeq
\beq
{d\delta v_R\over dt} - 2\Omega \delta v_\phi + i k_R\left(
{\delta P\over \rho} + {\bb {B\cdot\delta B}\over 4 \pi \rho}\right)
- i \left( \bb{k\cdot B}\over 4 \pi \rho\right) \delta B_R = 0,
\eeq
\beq
{d\delta v_Z\over dt} + i k_Z\left(
{\delta P\over \rho} + {\bb {B\cdot\delta B}\over 4 \pi \rho}\right)
- i \left( \bb{k\cdot B}\over 4 \pi \rho\right) \delta B_Z = 0,
\eeq
\beq
{d\delta v_R\over dt} +{\kappa^2\over  2\Omega} \delta v_R + i {m
\over R} \left(
{\delta P\over \rho} + {\bb {B\cdot\delta B}\over 4 \pi \rho}\right)
- i \left( \bb{k\cdot B}\over 4 \pi \rho\right) \delta B_\phi = 0.
\eeq

For the induction equations, it is helpful to define an
auxiliary vector $\bb{A}$ by
\beq \bb{A} = -\eta k^2 \bb{\delta B} + {c \bb{(k\cdot B)}\over 4\pi n_e e}
\bb{( k\times \delta B)}.
\eeq
The magnetic field equations are then
\beq
{d\delta B_R\over dt} = i \bb{(k\cdot B)} \delta v_R + A_R,
\eeq
\beq
{d\delta B_Z\over dt} = i \bb{(k\cdot B)} \delta v_Z + A_Z,
\eeq
\beq
{d\delta B_\phi\over dt} = i \bb{(k\cdot B)} \delta v_\phi +
{d\Omega\over d\ln R} \delta B_R+ A_\phi.
\eeq
The above equations imply that 
\beq
{d\bb{(k\cdot B)}\over dt} = - \eta k^2 \bb{ (k\cdot B)},
\eeq
so that if chosen to vanish initially, $\bb{k\cdot \delta B}$
will vanish throughout the evolution in a numerically stable
manner.  If $\bb{k\cdot\delta B}$ vanishes, then $\bb{k\cdot A} = 0$.

Because of the time dependence of the wavevector, the nonaxisymmetric
problem is more complex than its axisymmetric counterpart.
The approach we shall use is to reduce the system to two coupled second order
differential equations for $\delta B_R$ and $\delta B_\phi$,
which are then solved numerically (Balbus \& Hawley 1992b).  
This is an algebraically tedious process, and we present only the final result:

%\begin{eqnarray}
%{d^2\delta B_R\over dt^2} -{2m\over R}{k_R\over k^2} {d\Omega\over
%d\ln R}{d\delta B_R\over dt} + {2m\Omega k_Z\over Rk^2}{d\delta
%B_Z\over dt} +{2\Omega R\over m}
%\left( k_Z{d\delta B_Z\over dt} + k_R{d\delta B_R\over dt}\right){k_Z^2\over k^2}
%\nonumber\\
%+\bb{(k\cdot v_A)^2}+ 2\Omega A_\phi\left( k_Z^2 + m^2/R^2\over k^2 \right) - 
%{dA_R\over dt} + {2A_R m k_R\over Rk^2} {dR\Omega\over dR} = 0,
%\end{eqnarray}

\begin{eqnarray}
\frac{d}{dt} \left( \frac{d \delta B_r}{dt} -A_r \right) = 
-\frac{2r \Omega}{m} \frac{k_z^2}{k^2} 
\left[ k_r \left( \frac{d \delta B_r}{dt} -A_r \right) + 
k_z \left( \frac{d \delta B_z}{dt} -A_z \right) \right] \nonumber \\
+ \frac{2mk_r}{k^2} \frac{d \Omega}{dr} 
\left( \frac{d \delta B_r}{dt} -A_r \right) - 
\frac{2 m \Omega k_z}{r k^2} \left( \frac{d \delta B_z}{dt} -A_z \right) 
- \left( {\mathbf k} \cdot {\mathbf v}_A \right)^2 \delta B_r ,
\label{nax1}
\end{eqnarray}

%\begin{eqnarray}
%{d^2\delta B_Z\over dt^2} - {2m k_Z\over k^2 R} {d\Omega\over d\ln R}
%{d\delta B_R\over dt} -{2\Omega R \over m} \left( 1 + {m^2\over R^2
%k_R^2} \right)
%\left( k_R k_Z\over k^2\right) \left( k_Z { d \delta B_Z\over dt}
%+ k_R {d\delta B_R\over dt} \right) \nonumber\\
%+ {k_Z^2\over k_R k^2} {2\Omega m\over R} {d\delta B_Z\over dt}
%+ \bb{ (k\cdot v_A)}^2 \delta B_Z
%+{2mk_Z\over Rk^2} A_R {dR\Omega\over dR}
%-{dA_Z\over dt}
%-{2\Omega k_Rk_Z\over k^2}A_\phi =0.
%\end{eqnarray}

\begin{eqnarray}
\frac{d}{dt} \left( \frac{d \delta B_z}{dt} -A_z \right) = 
\frac{2r \Omega}{m} \frac{k_r k_z}{k^2}
\left[ k_z \left( \frac{d \delta B_z}{dt} -A_z \right) + 
k_r \left( \frac{d \delta B_r}{dt} -A_r \right) \right] \nonumber \\
+ \frac{2mk_z}{r k^2} \frac{d \left( r \Omega \right)}{dr}
\left( \frac{d \delta B_r}{dt} -A_r \right) 
- \left( {\mathbf k} \cdot {\mathbf v}_A \right)^2 \delta B_z .
\label{nax2}
\end{eqnarray}

\noindent These equations are formally the same as equations (2.19) and (2.20) of 
Balbus \& Hawley (1992b), but with the first-order time
derivatives of $\delta {B}$ transformed to

\begin{displaymath}
\frac{d \delta \bb {B}}{dt} \rightarrow \frac{d \delta \bb{
B}}{dt} - \bb{A} .
\end{displaymath}

\subsubsection{Behavior of Perturbations}

Although equations (\ref{nax1}) and (\ref{nax2}) appear rather opaque,
it is possible to understand the onset of instability by focusing on
the comparatively simple axisymmetric inequality (\ref{necc}).  In
essence, we find that the nonaxisymmetric evolution unfolds as a series
of axisymmetric problems.  In going from axisymmetry to nonaxisymmetry,
the only term in the ${\cal C}_0$ stability criterion that acquires a
time-dependence is $k$.  Consider the evolution of a strongly
leading disturbance, $k_R$ large and negative.  As $k_R$ initially moves
towards zero, the factor $(k_Z/k)^2$ goes from much less than, to
nearly unity (assuming $m^2/(k_Z R)^2$ is small).  The right hand side
of (\ref{necc}) starts out very small, and grows to its maximum value of
$|d\Omega^2/d\ln R|$.  At this stage, the formal instability criterion must be
satisfied, or it will never be satisfied over the entire course of the
evolution.  Then, as $k_R$ grows from zero to large and positive, the
right hand side of (\ref{necc}) once again diminishes with time, and
eventually the flow will stabilize.

\begin{figure}
\epsscale{0.7}
\plotone{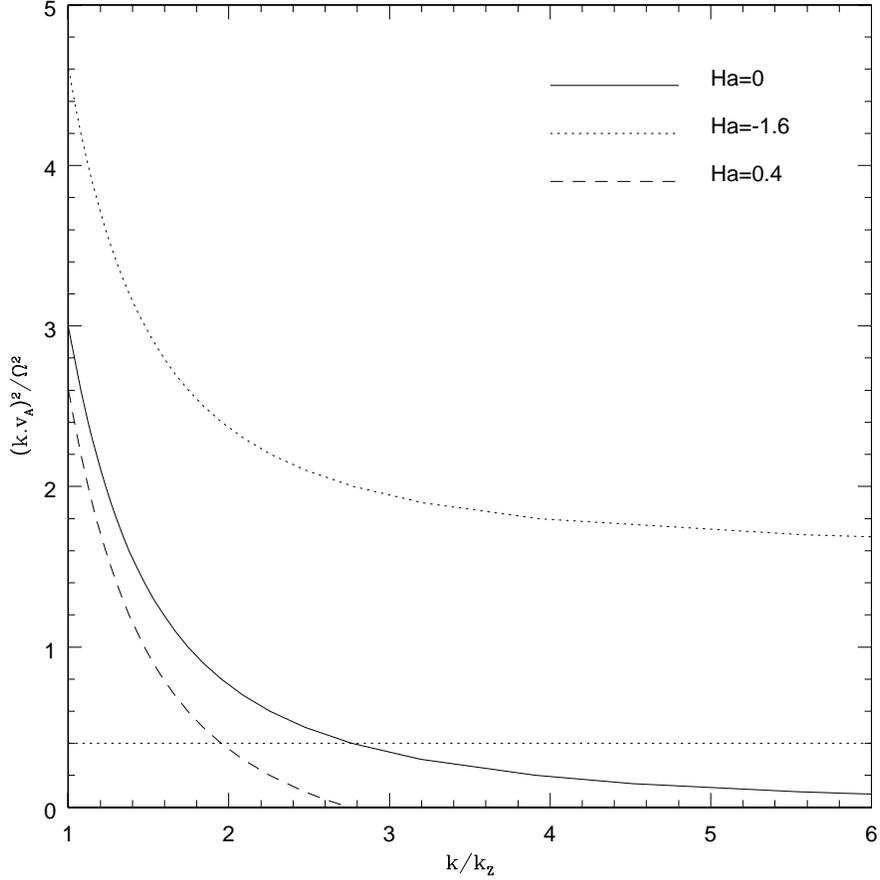}
\caption[]{Stability and instability regions in the $(\bb{k \cdot
v_A})^2/\Omega^2-(k/k_Z)$ plane for nonaxisymmetric disturbances, which
correspond as well to axisymmetry.  The Hall parameters $Ha$ are:
$Ha=0$ ({\em solid line}), $Ha=-1.6$ ({\em dotted line}) and $Ha=0.4$
({\em dashed line}).  For $Ha=0$ and $Ha=0.4$, the unstable region is
under the curve.  For $Ha=-1.6$, it is between the two curves.  $\bb{k
\cdot v_A}$ is constant for a shearing wavevector.  For strongly
leading disturbances, $k/k_Z$ is initially large, and the point
corresponding to the wavevector moves to the left in the plane on a
constant $\bb{k \cdot v_A}$ line.  The minimum value of $k/k_Z$ is
unity (to order $m^2/(k_Z^2 R^2$)).  After attaining its minimum, the
wavector point retraces its path to the right.  For values of $\bb{k
\cdot v_A}$ smaller than some critical value which depends on $Ha$, a
finite portion of time is spent in the unstable region, and substantial
growth may occur.}
\label{fig1}
\end{figure}

The situation is summarized in figure~(\ref{fig1}), which shows the
stability and instability regions of a Keplerian disk in the $(k/k_Z),
(\bb{k \cdot v_A}/\Omega)^2$ plane.  We have adopted the nominal
values $Ha=-1.6$ and $Ha=0.4$, and for comparison have also displayed
the magnetorotational instability results ($Ha=0$) of Balbus \& Hawley
(1992b).  Note that despite the complexity of the full system of
equations, in the end the HEMF simply slides the hyperbolic stability
curve up and down along the ordinate.  Since $\bb{k\cdot v_A}$ is
constant, the evolving path of a perturbation is along a horizontal
line in this plane, since only the abscissa coordinate changes with
time.  The nature of the unstable response is then determined entirely
by how long this horizontal trajectory remains inside the 
instability region.

\begin{figure}
\plotone{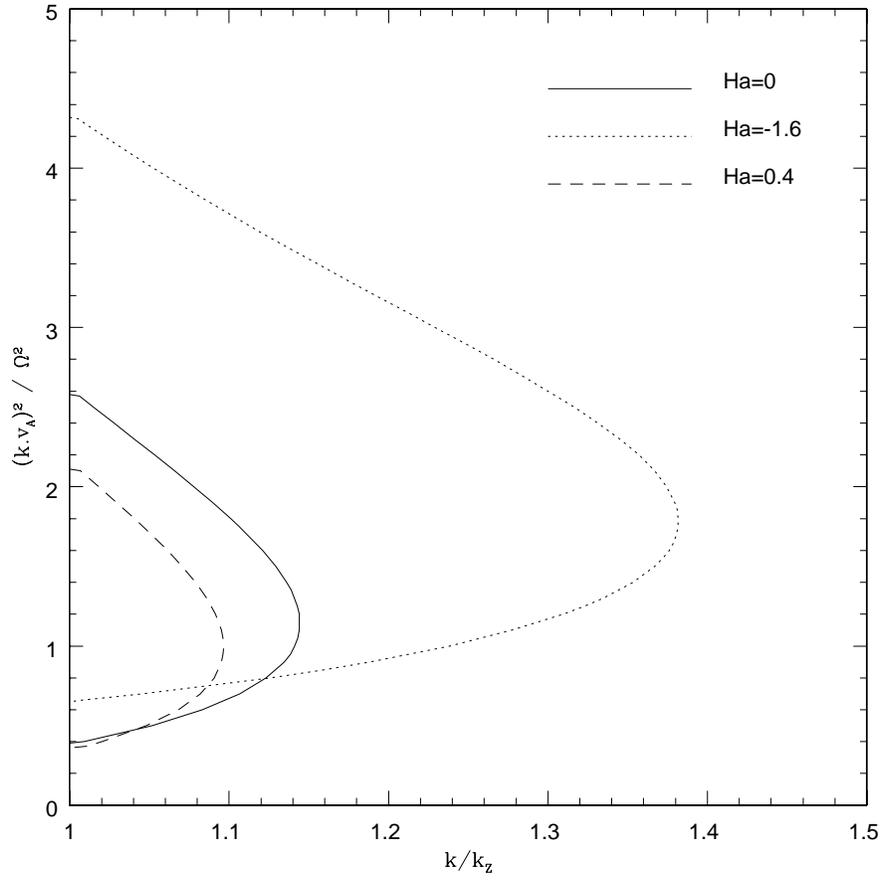}
\caption[]{Same as in Fig. \ref{fig1} but with a non zero resistivity
$\eta k_Z^2/\Omega=1$.  Here, for each value of $Ha$,  the unstable region
is between the curve and the vertical axis.  The resistivity
stabilizes the large $k_R$ disturbances.}
\label{fig2}
\end{figure}

\begin{figure}
\epsscale{0.5}
\plotone{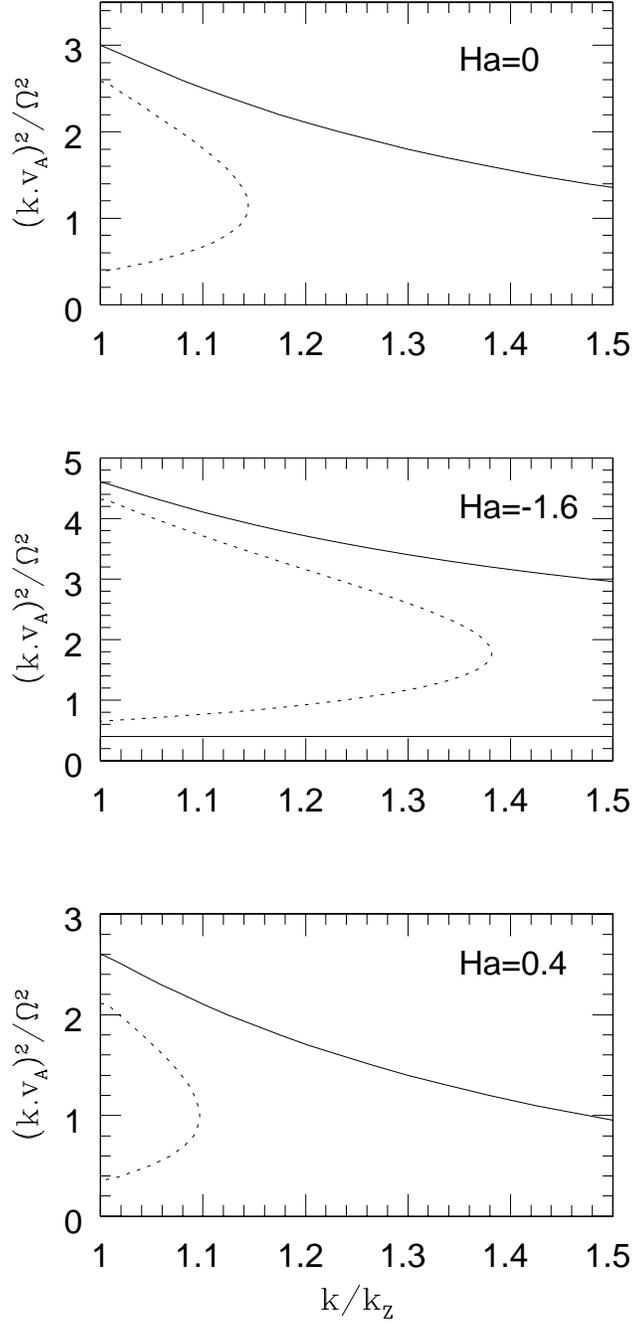}
\caption[]{Same as in Fig. \ref{fig1} and \ref{fig2} but here, to allow 
for comparison, we have plotted both the cases $\eta=0$ ({\em solid
line}) and $\eta k_Z^2/\Omega=1$ ({\em dotted line}) on the same
panel.  The Hall parameters are as shown.}
\label{fig3}
\end{figure}

In figure~(\ref{fig2}), both the Hall effect and resistivity are
included.  From equations (\ref{inst1}) and (\ref{inst2}), we see that
the marginal stability curves in the $(k/k_Z), (\bb{k\cdot
v_A}/\Omega)^2$ plane depend on the resistivity only through the
parameter $\eta k_Z^2/\Omega$.   We have adopted $\eta
k_Z^2/\Omega=1$.   To orient oneself, this would correspond to a disk
with aspect ratio $h/R=0.1$, $k_ZR =100$, and $(c_S/v_A) Re_M = 100$.
We see clearly in figure~(\ref{fig2}) that the resistivity stabilizes
small radial wavelengths.  In figure~(\ref{fig3}), to allow for direct
comparison, we have plotted both the resistive and nonresistive cases.

Figures~(\ref{fig1})--(\ref{fig3}) have been obtained strictly by
numerically solving equations~(\ref{nax1}) and (\ref{nax2}).
Nevertheless, each of these curves is almost indistinguishable from
those obtained simply by setting ${\cal C}_0=0$, where ${\cal C}_0$ is
defined by (\ref{inst1}).  Therefore, although the criterion
(\ref{inst2}) has been derived for axisymmetric perturbations, it is
very accurate for nonaxisymmetric disturbances also, with $k/k_Z$
varying with time.  Note that the curves displayed in
figures~(\ref{fig1})--(\ref{fig3}) depend on $m$ only implicitly,
through the ratio $k/k_Z$.  The role of $m$ is limited to determining
the evolution of $k/k_Z$, not the properties of the stability curves.

\begin{figure}
\epsscale {1.0}
\plotone{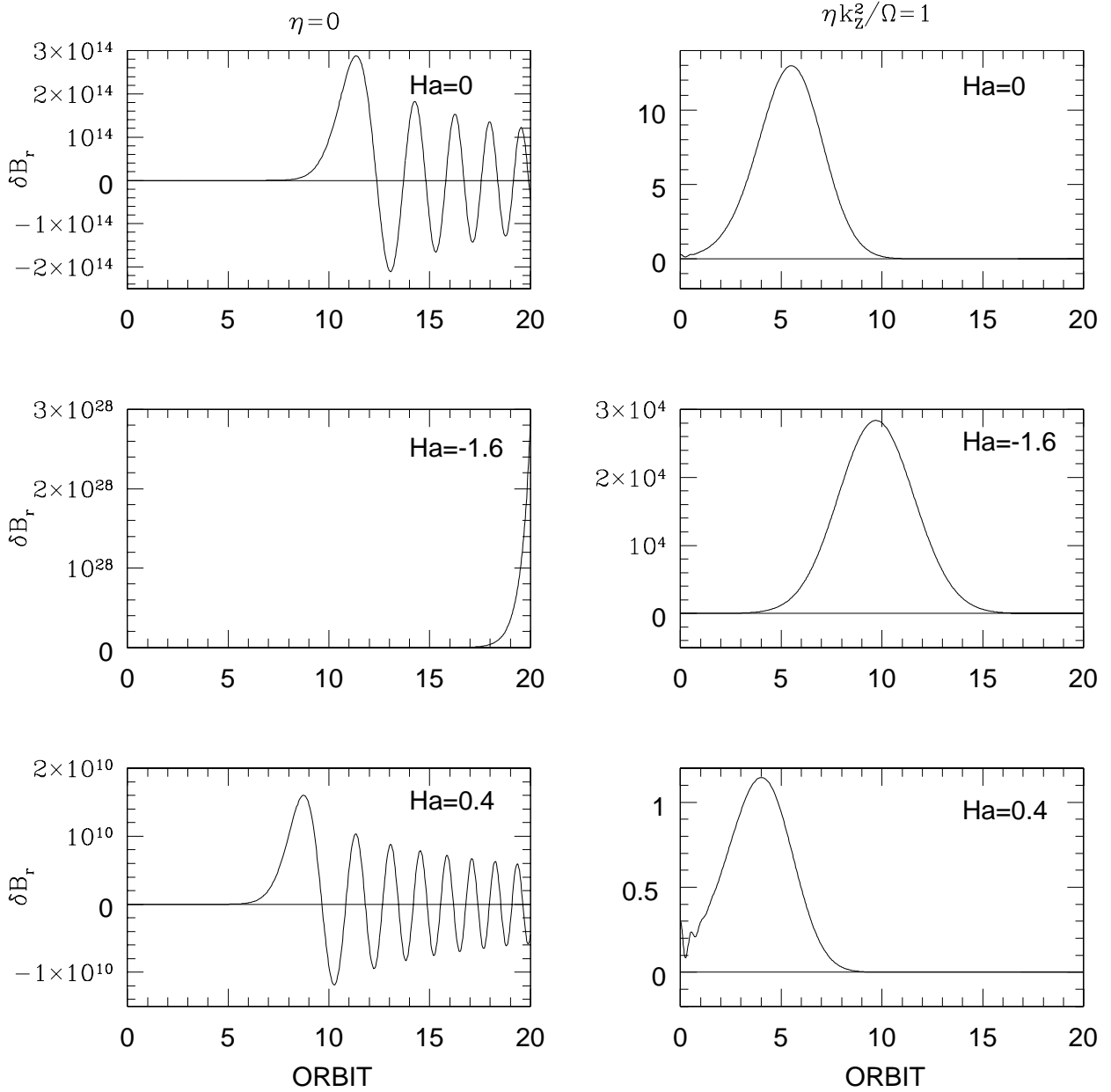}
\caption[]{Evolution of the perturbed radial field $\delta B_R$ for a
Keplerian disk with $(\bb{k \cdot v_A})^2/\Omega^2=1.5$.  The initial
amplitudes are $\delta B_R=0.3$ and $\delta B_Z = \delta B_{\phi} =0$,
and the wavenumbers are $k_Z R=100$, $m=1$.   The initial value of $k_R
R$ is determined by $\del \cdot \delta \bb{B} =0$.  The left panels 
have zero resitivity, the right panels have $\eta k^2/\Omega = 1$.
Hall parameters are as shown.} 
\label{fig4}
\end{figure}

In figure~(\ref{fig4}), we show the evolution of the perturbed radial
field $\delta B_R$ (the curves for $\delta B_Z$ and $\delta B_{\phi}$
are similar) for the same $Ha$ values as above, and for both $\eta=0$
and $\eta k_Z^2/\Omega=1$.  The value of $(\bb{k\cdot
v_A})^2/\Omega^2=1.5$ is chosen  because it lies in the unstable regime
for an axisymmetric $\eta = 0$ disk, for all the Hall parameters.
Initially $k/k_Z$ is very nearly unity, and the perturbation grows
until the wavevector moves out of the relevant unstable region in the
$(k/k_Z), (\bb{k \cdot v_A}/\Omega)^2$ plane.  When $\eta=0$,
explosive growth is achieved for the three values of $Ha$ considered.
As expected, the cases $Ha=-1.6$ and $Ha=0.4$ are respectively more and
less unstable than the case $Ha=0$.  The effect of the resistivity is
to suppress the growth of the perturbation at large wavenumbers,
corresponding to very early and very late times in the evolution.  The
dramatic reduction of peak amplitudes is evident.  The $Ha<0$ case,
however, remains quite unstable even with $\eta k_Z^2/\Omega = 1$.

To conclude, we have found that if $m/k_Z R$ is small, the
nonaxisymmetric behavior of linear perturbations may be understood on
the basis of axisymmetric behavior.  (Large $m$ disturbances are
stable.)  Whether growth occurs or not depends upon the instantaneous
location of the wavenumber in the $(k/k_Z), (\bb{k \cdot
v_A}/\Omega)^2$ plane.  Shear causes the wavenumber to retrace a
horizontal line in this plane, first from right to left (for an
initially leading distrubance), then the reverse.  If this path takes
the wavenumber into an {\em axisymmetrically} unstable zone, the
disturbance grows.  The effect of resistivity, not surprisingly, is to
dampen growth, with the smallest wavelengths being the most affected.

\section {Summary}

In a gas which is at least partially ionized, the magnetic field is,
but for resistivity, frozen into the electron fluid.  The difference in
the mean electron velocity and the center-of-mass fluid velocity gives
rise to HEMFs in the gas.  Somewhat surprisingly, in astrophysical
environments these can often be more important than Ohmic dissipation,
even in a fully ionized plasma.  This point is often not appreciated.
In protostellar disks, the Hall effect is small compared with
resistivity only if the {\em neutral} density is much in excess of
$10^{18}$ cm$^{3}$, or if $v_A \ll c_S$.

The dynamical consequence of HEMFs is the appearance of whistler
waves as a mode of the gas response.  These waves are carried nominally
by the electrons, but inductive electron-ion coupling and ion-neutral
collisional coupling together assure that the bulk of the fluid is
involved.  It is the interplay between these right-handed circularly
polarized whistlers and the Coriolis driven epicycles that affects the
stability of magnetized disks.  Whistler waves become dynamically
important when the ionization fraction drops below
$\omega_{c\mu}/\Omega$.

The Hall effect allows disks with either decreasing outward or increasing
outward angular velocity profiles to become unstable.  By way of contrast,
the standard MRI affects only those disks
with a decreasing outward profile, even though any angular velocity gradient 
increases the rotation energy for fixed angular momentum. 
The maximum growth rate is always found to be the Oort A value,
$(1/2)|d\Omega/d\ln R|$.  

Whether HEMFs stabilize or destabilize a disk depends not upon
$\bb{\Omega\cdot B}$, but upon $\bb{(k\cdot\Omega)(k\cdot B)}$.
The latter should be negative for destabilization.  
If $\bb{B}$ has a radial component, then it is always possible
to find a wavenumber to make this parameter whatever value is desired.
Thus, Hall physics will always tend to make some wavenumbers
more unstable.  Whether or not the disk actually become unstable will
depend on the value of the resistivity.

The nonlinear consequences of HEMFs are likely to be particularly
important for our understanding of transitions between active and
quiescent states in both dwarf novae and FU Orionis outbursts.  Neither
of these systems is presently understood at a fundamental level.
Numerical simulations which include only Ohmic losses have yet to
demonstrate that MHD turbulence can be sustained at the ionization
fractions that may be present in these cool disks (Fleming et al.~2000,
Menou 2000).   Numerical simulations including both the Hall effect and
Ohmic dissipation should prove most informative.

\section*{Acknowledgements}

We thank J.~Hawley, K.~Menou, and J.~Stone for useful discussions, and
W.~Winters for his skillful preparation of several figures.  SAB is
grateful to the Institut d'Astrophysique de Paris for its hospitality
and visitor support.  Both authors would like to thank Prof.
J.~Papaloizou of the Astronomy Unit at QMW, where this work was
initiated, for his generous support under PPARC grant
PPA/V/O/1997/00261.  SAB is supported by NASA grants NAG 5--7500, NAG
5--9266, and NSF grant AST--0070979.

\end{document}